\documentstyle[12pt]{article}
\normalbaselines 
\def\be{\begin{equation}}
\def\ee{\end{equation}}
\begin{document}

\begin{center} 
\vspace*{1.0cm} 
{\LARGE {\bf Hadamard renormalization, conformal anomaly and 
cosmological 
event horizons\footnote{To appear in Phys. Rev. D 56, 1 (1997)}\\ }}  

\vskip 1.5cm 
{\large {\bf H. Ghafarnejad\\}} 

\vskip 0.5   cm
Physics Department, Sharif University of Technology, Tehran, Iran\\
\vskip 0.5   cm

{\large {\bf H. Salehi\footnote { e-mail address: 
salehi@netware2.ipm.ac.ir}}}

\vskip 0.5 cm

Institute for Studies in Theoretical Physics and
Mathematics, \\ P.O.Box 19395-1795, Niavaran-Tehran, Iran
\\ and\\  
Arnold Sommerfeld Institute for Mathematical Physics, TU Clausthal, \\
Leibnizstr. 10, D-38678 Clausthal-Zellerfeld, Federal
Republic of Germany\\ 
\vskip 0.5 cm

\end{center}

\vspace{1cm}

\begin{abstract}
The Hadamard renormalization prescription is used to derive
a two dimensional analog of the renormalized
stress tensor for a minimally coupled scalar field in 
Schwarzschild-de Sitter 
space time. In the two dimensional analog the minimal coupling
reduces to the conformal coupling and  
the stress tensor
is found to be determined by the (nonlocal) contribution of 
the anomalous trace and some additional parameters in close relation to 
the work [1].
To properly relate the stress tensor to the state of outwards signals coming
from the direction of the black hole at late times we propose a cut-off
hypothesis which excludes the contribution of the anomalous trace close
to the black hole horizon. The corresponding cut-off scale is found to be 
related to the equilibrium-temperature of the cosmological horizon
in a leading order estimate. 
Finally, we establish a relation between
the radiation-temperature of the black hole horizon at 
large distance from the
hole and the
the anomalous trace and determine the correction term to the Hawking
temperature due to the presence of the cosmological horizon. 

\end{abstract} 
\vspace{1cm}
\section{Introduction}
A central problem in quantum field theory in curved space time is 
the computation of the renormalized expectation value of the stress
tensor operator [2]. Usually one is inclined to expect that 
the stress tensor at some point in a curved space-time can be measured 
by a well defined local operator. However, the usual 
expression for the stress tensor operator involves singular products
of the field operator at the same space time point, and it seems clear
that such singular products do not allow the definition of a 
well-defined local operator. Renormalization theory of the stress tensor 
was originally designed to solve this problem. But, it must be remarked 
that the usual scheme of renormalization involves complicated, 
often ambiguous, steps and it is by no means apparent 
that the resulting final expressions 
actually correspond to the expectation value of a well-defined local
operator acting
on the Hilbert space of states. 
In principle, one should recognize that
there is no conceptual support for a local measure of energy momentum 
of some given state without reference to any global construct. In fact, 
even in Minkowski space energy momentum is measured relative to a global
construct, namely the Minkowski vacuum. We emphasize that the conceptual
basis of the renormalization theory, as it is currently understood, is still
ill defined.\\ 
Despite these difficulties, the usual renormalization prescriptions 
have some power, in that they satisfy some general requirements, such as
the covariant conservation law, and in the case of the 
conformal invariant coupling the general requirement on the anomalous trace.
It is just for this reason that the study of the usual renormalization
prescriptions can still be justified.
In this article we first clarify this aspect in connection with the Hadamard
renormalization prescription developed in [3][4][5], (see also [6]).
We then apply the results to the two dimensional analog 
of Schwarzschild-de Sitter space-time
and derive the leading order approximation of equilibrium temperature and
radiation temperature associated with the
cosmological event horizon and the black hole horizon respectively. 
In our presentation the equilibrium temperature 
of the cosmological event horizon basically emerges from a 
physical cut-off which close to the black hole 
horizon excludes the contribution
of the anomalous trace to radiation part of the stress tensor. 
In essence, such an approach has a certain similarity 
to the recent publications
[7][8][9] in which dissatisfactions were expressed with the use 
of the singular modes escaping from a region close to the black hole 
horizon in the
derivation of the Hawking effect. We believe that our presentation, 
although it will not
use the notion of mode to generate the Hawking effect, 
is instructive because it suggests that
in a future theory 
the physical cut-off for a black-hole may have an intimate connection
to the presence of an associated cosmological horizon.

\section{Hadamard renormalization}

We consider a linear scalar quantum field $\phi$ propagating 
on a curved space time with the 
action of the standard form [2]
\be
S[\phi]=-{1\over2}\int{d^4x g^{1/2}(g_{\alpha\beta}\phi^{;\alpha}
\phi^{;\beta}+\xi R\phi^2+m^2\phi^2)}.
\ee
where $m$ and $\xi$ are parameters, and $R$ is the 
scalar curvature (In the following the semicolon and $\nabla$ 
indicates covariant differentiation).
The corresponding field equation is
\be
(\Box-m^2-\xi R)\phi(x)=0.
\ee
The choice of the parameter $m$ and $\xi$ depends on the particular type
of coupling we wish to consider. For example, the minimal coupling
corresponds to ($m=0,\xi=0$) and the conformal coupling in four dimensions
corresponds to ($m=0,\xi=\frac{1}{6}$).\\
The energy momentum of $\phi$ is defined by the singular expression
\be
T^{\mu\nu}(x)=({1-2\xi}){\nabla^\mu}\phi{\nabla^\nu}\phi+({2\xi-{1\over2}})
g^{\mu\nu}{\nabla_\beta}\phi{\nabla^\beta}\phi+\xi(R^{\mu\nu}
-{1\over2}g^{\mu\nu}R)\phi^2
\ee
$$+2\xi\phi(g^{\mu\nu}\Box\phi-\nabla^\mu\nabla^\nu\phi)-
{1\over2}m^2g^{\mu\nu}\phi^2.$$
We shall deal with a particularly useful version of (3) in terms of 
anticommutator, namely
\be
T^{\mu\nu}(x)={1\over2}({1-2\xi})\{\nabla^{\mu}\phi,\nabla^{\nu}\phi\}+
({\xi-{1\over4}})
g^{\mu\nu}\{\nabla^{\beta}\phi,\nabla_{\beta}\phi\}
\ee
$$+{1\over2}\xi(R^{\mu\nu}
-{1\over2}g^{\mu\nu}R)\{\phi,\phi\}
+\xi g^{\mu\nu}\{\phi,\Box\phi\}-\xi\{\phi,\nabla^{\mu}
\nabla^{\nu}\phi\}
-{1\over4}m^2g^{\mu\nu}\{\phi,\phi\}.$$
A state of $\phi$ is characterized by a hierarchy of Wightman functions
\be
<\phi(x_1),...,\phi(x_n)>.
\ee
The 'operator' $T^{\mu\nu}$ takes a singular 
expectation value $<T^{\mu\nu}>$ in a given state. Using the point-splitting
method [10], this singular
expectation value can 
most
conveniently be represented by
\be
<T^{\mu\nu}>=lim_{x^\prime \to x}D^{\mu\nu}(x,x^\prime)\{G^+(x,x^\prime)\}.
\ee
Here $G^+(x,x^\prime)$ is the symmetric two-point function, 
$D^{\mu\nu}(x,x^{\prime})$ is the bilocal differential operator
\be
D^{\mu\nu}(x,x^\prime)=({{1\over2}-\xi})\{{g_{\mu^\prime}^\mu}
\nabla^{\mu^\prime}
\nabla^\nu
+{g_{\nu^\prime}^{\nu}}\nabla^\mu\nabla^{\nu^\prime}\}+({2\xi
-{1\over2}})g^{\mu\nu}g_{\beta^\prime}^{\beta}
\nabla_\beta\nabla^{\beta^\prime}+
\ee
$$\xi({R^{\mu\nu}-{1\over2}g^{\mu\nu}R})+
\xi g^{\mu\nu}\{\Box+\Box^{\prime}\}-\xi\{\nabla^{\mu}\nabla^{\nu}
+{g_{\mu^\prime}^\mu g_{\nu^\prime}^\nu}\nabla^{\mu\prime}
\nabla^{\nu\prime}\}
-{1\over2}m^2g^{\mu\nu}$$
and $g^\beta_{\beta^\prime}$ is the bivector of paralel transport.
This expression makes explicit that the
singular character of the operator $T^{\mu\nu}$ emerges as a consequence of 
the short-distance singularity of the symmetric two-point function 
$G^{+}(x,x^{\prime})$. This function satisfies the equation (2) in each
argument.\\
We remark that for a linear theory  the antisymmetric part of the two-point
function is common to all states in the same representation. It is 
just the universal commutator function. Thus, in our case 
all the relevant informations
about the state-dependent part of the two-point function are encoded in 
$G^{+}(x,x^{\prime})$.
Equivalence principle suggests that the leading singularity
of $G^{+}(x,x^{\prime})$ should have a close correspondence to
singularity structure of the two-point function
of a free massless field in Minkowski space [11]. 
In general the entire singularity of $G^{+}(x,x^{\prime})$ may have a more
complicated structure. Usually one assumes that $G^{+}(x,x^{\prime})$ has
a singular structure represented by the Hadamard expansions. This
means that in a normal neighbourhood of a point $x$ the function
$G^{+}(x,x^{\prime})$ can be written
\be
G^+(x,x^\prime)={1\over{8\pi^2}}\Biggl\lbrace{{{\Delta^{1/2}(x,x^\prime)}
\over{\sigma(x,x^\prime)}}+V(x,x^\prime) \ln{\sigma(x,x^\prime)}+
W(x,x^\prime)}\Biggr\rbrace
\ee
where $2\sigma(x,x^{\prime})$ is the square of the distance along the
geodesic joining $x$ and $x^{\prime}$ and $\Delta(x,x^{\prime})$ is 
the Van vleck determinant
\be
\cases{\Delta(x,x^\prime)=-g^{-1/2}(x)
Det\{-\sigma_{;\mu\nu\prime}\}g^{-1/2}(x^\prime)\cr
                    g(x)=Detg_{\alpha\beta}\cr}.
\ee
The functions $V(x,x^\prime)$ and $W(x,x^\prime)$ have the following 
representations as power series
\be
V(x,x^{\prime})=\sum_{n=0}^{+\infty}V_n(x,x^\prime)\sigma^n 
\ee
\be
W(x,x^{\prime})=\sum_{n=0}^{+\infty}W_n(x,x^\prime)\sigma^n
\ee
in which the coefficients are determined by applying the equation (2) 
to $G^{+}(x,x^{\prime})$, yielding the recursion relations
\be
(n+1)(n+2)V_{n+1}+(n+1)V_{n+1;\alpha}\sigma^{;\alpha}-
(n+1)V_{n+1}\Delta^{-1/2}\Delta^{1/2}_{;\alpha}\sigma^{;\alpha}+
\ee
$${1\over2}(\Box-m^2-\xi R)V_n=0$$
\be
(n+1)(n+2)W_{n+1}+(n+1)W_{n+1;\alpha}
\sigma^{;\alpha}-(n+1)W_{n+1}\Delta^{-1/2}\Delta^{1/2}_{;\alpha}
\sigma^{;\alpha}+
\ee
$${1\over2}(\Box-m^2-\xi R)W_n+(2n+3)V_{n+1}+V_{n+1;\alpha}
\sigma^{;\alpha}-V_{n+1}\Delta^{-1/2}\Delta^{1/2}_{;\alpha}
\sigma^{;\alpha}=0$$
together with the boundary condition
\be
V_0+V_{0;\alpha}\sigma^{;\alpha}-V_0\Delta^{-1/2}
\Delta^{1/2}_{;\alpha}\sigma^{;\alpha}+
{1\over2}(\Box-m^2-\xi R)\Delta^{1/2}=0.
\ee
From these relations one can determine the function $V(x,x^{\prime})$
uniquely in terms of local geometry. It takes therefore the same universal
form for all states. But the biscalar $W_{0}(x,x^{\prime})$ remaines
arbitrary. Its specification depends significantly on the choice of a 
state.\\Let us now explain the standard Hadamard renormalization
prescription.
The basic strategy is, in the first place, to extract the finite part of
$G^{+}(x,x^{\prime})$ by subtracting from $G^{+}(x,x^{\prime})$ a local 
symmetric two-point function $G^+_L(x,x^{\prime})$ with the same 
short-distance singularity of the
Hadamard expansion and, in the second place, to define the renormalized 
expectation value of the stress
tensor as
\be
<T^{\mu\nu}>_{ren}= lim_{x^{\prime}\rightarrow x}
D^{\mu\nu}(x,x^{\prime})\{G^{+}(x,x^{\prime})-G^{+}_{L}(x,x^{\prime})\}.
\ee
The result is apparently finite. But there is a fundamental ambiguity
concerning the choice of $G^{+}_{L}(x,x^{\prime})$. As a general criterion
one reasonably assumes that $G^{+}_{L}(x,x^{\prime})$ is a function of
local geometry.\\
This criterion does not eliminate the ambiguity concerning 
the choice of the function $G^{+}_{L}(x,x^{\prime})$, but 
the renormalization theory replaces this ambiguity
by another one, namely 
the freedom to add   
to $<T^{\mu\nu}>_{ren}$ a state-independent conserved tensor. 
We explain the underlying reasoning. Using the definitions (7),(8) and (15), 
one can write  the decomposition
\be
<T^{\mu\nu}>_{ren}= lim_{x^\prime\rightarrow x}
D^{\mu\nu}(x,x^{\prime})\{{(8\pi^2)^{-1}}W(x,x^{\prime})\}+ \Sigma^{\mu\nu}
\ee
in which the first term on the righthand side represents
the finite state-dependent contribution of the function $W(x,x^{\prime})$
in the Hadamard expansion of $G^+(x,x^\prime)$, 
and the second term is imagined to incorporate the finite 
state-independent contribution 
of $G^{+}(x,x^{\prime})$ together with the finite 
state-independent contribution of $G^{+}_{L}(x,x^{\prime})$. Now the point is that
the conservation
law determines the tensor $\Sigma^{\mu\nu}$ up to a divergence-less 
state-independent tensor. Thus the ambiguity concerning the choice 
of $G^{+}_{L}(x,x^{\prime})$ yields the freedom to add to the tensor 
$\Sigma^{\mu\nu}$ a conserved state-independent tensor.\\
The decomposition (16) is, however, incomplete without
specifying the nature of the tensor $\Sigma^{\mu\nu}$. In the  
renormalization theory one uses a decomposition in which the tensor 
$\Sigma^{\mu\nu}$ comes out to be divergence-less. 
To find the corresponding decomposition
we apply the conservation law to $<T^{\mu\nu}>_{ren}$ and find
for the divergence of $\Sigma^{\mu\nu}$ the expression
\be
\nabla_{\mu} \Sigma^{\mu\nu}=-\nabla_{\mu} \Gamma^{\mu\nu}[W]
\ee
where 
\be
\Gamma^{\mu\nu}[W]= lim_{x^{\prime}\rightarrow x}
D^{\mu\nu}(x,x^{\prime}) \{{(8\pi^2)^{-1}}W(x,x^{\prime})\}.
\ee
Now expanding 
$W(x,x^\prime)$ into a covariant power series [6][10][12], namely
\be
W(x,x^\prime)=W(x)-{1\over2}W_{;\alpha}(x)\sigma^\alpha
+{1\over2}W_{\alpha\beta}(x)\sigma^\alpha\sigma^\beta+
{1\over4}\{{1\over6}W_{;\alpha\beta\gamma}
-W_{\alpha\beta;\gamma}\}\sigma^\alpha\sigma^\beta\sigma^\gamma+O(\sigma^2),
\ee
the tensor $\Gamma^{\mu\nu}[W]$ can be calculated to yield 
\be
\Gamma^{\mu\nu}[W]={(8\pi^2)^{-1}}\{{1\over2}({1-2\xi})W^{;\mu\nu}(x)
+{1\over2}({2\xi-{1\over2}})\Box W(x)g^{\mu\nu}+
\ee
$$\xi({R^{\mu\nu}-{1\over2}g^{\mu\nu}R})W(x)-
{1\over2}m^2g^{\mu\nu}W(x)-
({W^{\mu\nu}-{1\over2}g^{\mu\nu}W^\gamma_\gamma})\}.$$
In this expression the scalar $W(x)$ is arbitrary, but once its has been
chosen the choice of the tensor $W_{\mu\nu}(x)$ must be subjected to the
the constraint
\be
\nabla_\mu (W^{\mu\nu}-{1\over2}g^{\mu\nu}(W_\gamma^\gamma-m^2W+{1\over2}
\Box W))
=2v_1^{;\nu}+{1\over2}R^{\mu\nu}W_{;\mu}-{1\over2}\xi RW^{;\nu}
\ee
which, using the differential identity 
\be
\Box(\nabla^{\nu})W=\nabla^{\nu}\Box W+R^{\mu\nu}\nabla_{\mu}W,
\ee
follows from the symmetry property of the biscalar 
$W(x,x^\prime)$ together with (see, [6][12]) 
\be
(\Box-m^2-\xi R)W(x,x^\prime)=-6v_1(x)+
2v_{1;\alpha}\sigma^{;\alpha}+O(\sigma).
\ee
From the constraint (21) one finds  
\be
\cases{\nabla_{\mu} \Sigma^{\mu\nu}={2(8\pi^2)^{-1}}
\nabla_{\mu}g^{\mu\nu}v_1(x)\cr
lim_{x^\prime \to x}V_1(x,x^\prime)=v_1(x)\cr}.
\ee
This has the implication that 
the incorporation of a compensating term proportional to 
$g^{\mu\nu}v_1(x)$ into the tensor
$\Gamma^{\mu\nu}[W]$ will make $\Sigma^{\mu\nu}$ divergence-less.
The corresponding decomposition used by the renormalization theory is 
\be
<T^{\mu\nu}>_{ren}= \bar{\Gamma}^{\mu\nu}[W]+\Sigma^{\mu\nu}
\ee
where
\be
\bar{\Gamma}^{\mu\nu}[W]=\Gamma^{\mu\nu}[W]+{2(8\pi^2)^{-1}}g^{\mu\nu} v_1(x).
\ee
The merits of such a decomposition is that each term becomes now 
divergence-less.\\
For the calculations, the tensor $\bar{\Gamma}^{\mu\nu}[W]$ is very important
because, firstly, it provides a conserved tensor which contains  
all the relevant informations about the state-dependent
part of $<T^{\mu\nu}>_{ren}$ and, secondly, in the case of conformal
coupling it produces the usual restriction imposed on the anomalous trace of
$<T^{\mu\nu}>_{ren}$ [1][13]. In the following we shall exclusively deal
with the tensor $\bar{\Gamma}^{\mu\nu}[W]$.
From (20) and (26) one gets its explicit
expression as
\be
\bar{\Gamma}^{\mu\nu}[W]=(8\pi^2)^{-1}\{{1\over2}({1-2\xi})
W^{;\mu\nu}(x)+{1\over2}({2\xi-{1\over2}})\Box W(x)
g^{\mu\nu}+\xi(R^{\mu\nu}-{1\over2}g^{\mu\nu}R)W(x)-\qquad
\ee
$${1\over2}m^2g^{\mu\nu}W(x)-(W^{\mu\nu}-{1\over2}g^{\mu\nu}W_\gamma^\gamma)
+2g^{\mu\nu}v_1(x)\}.$$

\section{The approximate stress tensor in the presence of 
the cosmological constant}
  
We study now the case of minimal coupling $\xi=m=0$, and 
proceed to find the approximate form of 
the tensor $\bar{\Gamma}^{\mu\nu}[W]$ in 
a space-time with a metric given by a vacuum solution of 
Einstein's equations in the presence of a small  
cosmological constant $\Lambda$. The metric arises as a solution 
of the equations  
\be
G^{\mu\nu}+\Lambda g^{\mu\nu}=0.
\ee
from which One finds
\be
\cases{R^{\mu\nu}=O(\Lambda)\cr
R^{\alpha\beta\gamma\delta}= O(\Lambda)\cr
R=O(\Lambda)\cr}.                
\ee
where $O(\Lambda)$ indicates the order of the tensors involved 
with respect to the cosmological constant [ see appendix.~A].
We now consider the construction of the tensor $W^{\mu\nu}$ for 
the particularly simple case in which the
scalar $W(x)$ is taken as slowly varying space-time
function. In this case the function $W(x)$ can approximately be replaced 
by an almost constant mean value $\bar{W}$, so we can neglect its
derivatives. Correspondingly, the divergence relation (21) results in the
following constraint
\be
\nabla_{\mu}(W^{\mu\nu}-\frac{1}{2}g^{\mu\nu}
W_{\alpha}^{\alpha})=O(\Lambda^{2}).
\ee
Using the relations (23) one can obtain a 
further constraint on the trace of the
tensor $W^{\mu\nu}$. One finds
\be
W_{\alpha}^{\alpha}=O(\Lambda^{2})
\ee
which together with (30) implies 
\be
\nabla_{\mu}W^{\mu\nu}=O(\Lambda^{2}).
\ee                   
Thus, up to terms of order $\Lambda^{2}$, 
the construction of the tensor
$\bar{\Gamma}^{\mu\nu}[W]$  amounts to finding the traceless conserved
tensor $W^{\mu\nu}$. For our purpose it is convenient to use 
for $W^{\mu\nu}$ a decomposition of the form
\be
W^{\mu\nu}=-\alpha\bar{W}G^{\mu\nu}-S^{\mu\nu}
\ee
where $\alpha$ is a constant parameter. 
As a consequence of Bianchi identity $\nabla_{\mu}G^{\mu\nu}=0$ one gets 
then from (30)-(32) the corresponding
constraints on the tensor $S^{\mu\nu}$, namely
\be
\nabla_{\mu}S^{\mu\nu}=O(\Lambda^{2}),~~~S_{\alpha}^{\alpha}=\alpha\bar{W}R+
O(\Lambda^{2}).
\ee
Our approximation now consists in neglecting terms of 
order $\Lambda^{2}$. One finds from
(27), (33) and (34) the approximate expression 
of the tensor $\bar{\Gamma}^{\mu\nu}[W]$ in terms of $S^{\mu\nu}$, namely 
\be
{\bar{\Gamma}}^{\mu\nu}[W]\approx S^{\mu\nu}+\alpha\bar W G^{\mu\nu}
\ee
where the tensor 
$S^{\mu\nu}$ is a conserved tensor with the trace  
$S_{\alpha}^{\alpha}=\alpha\bar{W}R$.

\section{Dimensional reduction}

Our goal now is to arrive at a (suitably defined) 
two dimensional analog of the approximate 
stress tensor (35). In two dimensions the field $\phi$ is a dimensionless
quantity. Correspondingly, the stress tensor takes the dimension
of a length to the power $-2$. Thus, to arrive at a two dimensional analog 
of the stress tensor we first replace the field $\phi$ by the
dimensionless quantity $\bar W^{-1/2}\phi$. Correspondingly, we replace the
tensor $\bar{\Gamma}^{\mu\nu}[W]$ by 
$\bar W^{-1}\bar{\Gamma}^{\mu\nu}[W]$. Denoting this latter quantity
by $\bar{\Gamma}^{\mu\nu}_{2}[W]$
and taking into account that in two dimensions the tensor
$G^{\mu\nu}$ is identically vanishing, 
we define the two dimensional analog of (35) as 
\be
\cases{\bar{\Gamma}^{\mu\nu}_{2}[W]\approx 
S^{\mu\nu}_{2},\cr
S^{\mu\nu}_{2}=\bar W^{-1} S^{\mu\nu} \cr}
\ee
in which the conserved tensor $S_{2}^{\mu\nu}$ takes now the trace 
\be
S_{2~\alpha}^{\alpha}=\alpha R.
\ee
The still unknown parameter $\alpha$ in (37) 
can be determined by a general requirement.
We remark that the minimal coupling in two
dimension reduces to the conformal coupling. Thus, $\alpha$
can be determined by the requirement
that the trace of $S^{\mu\nu}_{2}$ shall reproduce the general
restriction on the anomalous trace in two dimensions [1][13], yielding
$\alpha={{1}\over{24\pi}}$. 
We conclude that, in our two dimensional analog of the problem, 
the determination                                              
of the tensor $\bar{\Gamma}^{\mu\nu}[W]$ amounts to finding
a tensor $S^{\mu\nu}$ satisfying the constraints 
(We suppress the subscript 2)
\be
\cases{\nabla_{\mu}S^{\mu\nu}=0,\cr
    S_{\alpha}^{\alpha}={{1}\over{24\pi}}R\cr}.
\ee
These constraints corresponds exactly to the well known constraints imposed
on the two dimensional stress tensor of a conformally invariant field. Here
we have shown that, restricting ourselves to solutions of (28), 
these constraints can also be found from a (suitably defined) dimensional
reduction of the state-dependent part of the renormalized stress tensor
of a minimally coupled field in four dimensions plus some approximation.

\section{Thermal Radiation and the cosmological event horizons}

As an illustration we shall apply the results of the previous sections
to a particular solution of the equations (28), 
namely the Schwarzschild-de Sitter space time 
on which the metric in the static 
and spherical symmetric 
form is given by 
\be
ds^{2}=-(1-\frac{2M}{r}-\frac{\Lambda r^{2}}{3}) dt^{2}+
(1-\frac{2M}{r}-\frac{\Lambda r^{2}}{3})^{-1} dr^{2}+r^{2}
(d\theta^{2}+sin^{2} \theta  d\phi^{2}).
\label{T1}\ee
This metric describes a Schwarzschild-like black hole in the presence 
of the cosmological constant $\Lambda$ [14].                  
In the following we shall restricts us to a typical situation in which 
$\Lambda>0$ and $\Lambda M^{2}<<1$. 
There are two solutions
of $g_{tt}=0$ corresponding to 
a black hole horizon and a cosmological horizon. The black hole horizon
can be obtained if we approximate $g_{tt}$ for $r-2M<<2M$ by
\be
g_{tt}\approx -(1-\frac{(2M+\Lambda(2M)^{3}/3}{r})
\label{d2}\ee
from which one infers that $g_{tt}$
becomes zero at a value $r_b\approx 2M(1+\frac{4}{3}\Lambda M^{2})$. 
This is the position of          
black hole horizon. It increases with respect to
the Schwarzschild-radius
$r=2M$ by a term of the relative order ${\Lambda\over3} (2M)^{2}$. 
The cosmological horizon can be obtained if we approximate $g_{tt}$ for 
$\sqrt{3\over\Lambda}-r<<\sqrt{3\over\Lambda}$ by
\be
g_{tt}\approx -({1-({{{2M}\over{(\sqrt{3/\Lambda})^3}}+{\Lambda\over3}})r^2})
\ee
from which one infers that $g_{tt}$ becomes zero at a value 
$r_c\approx\sqrt{3/\Lambda}(1-M{\sqrt{\Lambda\over 3}})$. 
This is the position of cosmological horizon. It decreases 
with respect to the de Sitter radius 
$r=\sqrt{3\over\Lambda}$ by a term of the relative order 
$\sqrt{\Lambda M^2/3 }.$\\
In the following we shall deal with the two dimensional analog of the metric
(39), namely
\be
ds^{2}=-(1-\frac{2M}{r}-\frac{\Lambda r^{2}}{3}) dt^{2}+
(1-\frac{2M}{r}-\frac{\Lambda r^{2}}{3})^{-1} dr^{2}
\label{T2}\ee
for which the positions of event horizons are the same
as those for the four dimensional case. 
The metric (42) can be written in the conformally-flat form
\be
ds^2=\Omega(r)(-dt^2+dr^{*2})\qquad\qquad
\ee
with
\be
\Omega(r)=1-{2M\over r}-{\Lambda r^2\over3},~~~{{dr}\over{dr^*}}=\Omega(r).
\ee
In the following our main objective is the determination of the tensor 
$S^\nu_\mu$ defined by (38) for the metric of (42).
For the nonzero christoffel symbols of the metric (42) we have in $(t,r^*)$ 
coordinates
\be
\Gamma^{r^*}_{tt}=
\Gamma^t_{tr^*}=
\Gamma^t_{r^*t}=\Gamma^{r^*}_{r^*r^*}={1\over2}{d\over{dr}}\Omega(r).
\ee
Under the assumptions that $S_{\mu}^{\nu}$ is time-independent and 
spherically 
symmetric, the conservation equation takes the form
\be
\partial_{r^*}S_t^{r^*}+\Gamma^t_{tr^*}S^{r^*}_t-
\Gamma_{tt}^{r^*}S^t_{r^*}=0
\ee
\be
\partial_{r^*}S^{r^*}_{r^*}+
\Gamma^t_{tr^*}S^{r^*}_{r^*}-\Gamma^t_{tr^*}S^t_t=0
\ee
with
\be
\cases{S^t_{r^*}=-S^{r^*}_t\cr
S^t_t=S^\alpha_\alpha-S^{r^*}_{r^*}\cr}
\ee
where $S^\alpha_\alpha$ is trace anomaly in two dimensions.
Using the equations $(45-48)$ one can show that
\be
{d\over{dr}}\{\Omega(r)S^{r^*}_t\}=0
\ee
and
\be
{d\over{dr}}\{\Omega(r)S^{r^*}_{r^*}\}=
{1\over2}\Bigl\lbrace{d\over{dr}}\Omega(r)\Bigr\rbrace S^\alpha_\alpha.
\ee
Equation $(49)$ leads
\be
S^{r^*}_t=\alpha\Omega^{-1}(r)
\ee
where $\alpha$ is a constant of integration. The solution of $(50)$ 
may be written in the following form 
\be
S^{r^*}_{r^*}(r)=(H(r)+\beta)\Omega^{-1}(r),~~~\beta=\Omega(L)
S^{r^*}_{r^*}(L)
\ee
where
\be
H(r)={1/2}\int^r_{L} 
S_\alpha^\alpha(r^\prime){d\over{dr^\prime}}\Omega(r^\prime)dr^\prime
\ee
with $L$ being an arbitrary scale of length, and
\be
S_\alpha^\alpha(r)=
{1\over{24\pi}}R={M\over{6\pi r^3}}+{1\over{36\pi}}\Lambda.
\ee
Given a length scale $L$, the function $H(r)$ incorporates the
corresponding (non-local) contribution of the trace $S_\alpha^\alpha(r)$
to the tensor $S_{\mu}^{\nu}$. The choice of $L$ needs careful considerations.
It does not appear possible to include the contribution of a region
very close to the black hole horizon to the off-diagonal components of 
$S_{\mu}^{\nu}$, if the latter is taken as properly describing 
the late time (steady state) behavior of outwards signals 
coming from
the direction of the black hole\footnote{Of course, from (52) it follows
that the function
$H(r)$ has no explicit contribution to 
the off-diagonal components of the stress tensor. But we shall see later,
equation (66), that $H(r)$ has an implicit "off-diagonal" contribution 
to the radiation temperature of the black hole through
the parameter $\alpha$ appearing in (51)}. 
In fact, 
a "off-diagonal" contribution of a region very close to the black hole 
horizon can not be sharply defined with respect to the state of 
outwards signals at 
late times
because the infinite gravitational redshift at the black
hole horizon connects the latter state at asymptotic times with
the physical situations in the vicinity of the horizon where the
quantum fluctuation of the horizon (and the corresponding change of
the gravitational field) can no longer 
be neglected.
To accurately describe the outwards signals at late time 
by $S_{\mu}^{\nu}$ our criterion is to exclude in 
the definition of $H(r)$ the contribution
of the trace very close to the black hole horizon using 
a characteristic cut-off length $l_{c}$.
Since the scales of 
the problem is set by the mass of the black hole and the cosmological
constant, it should be possible to define the cut-off in 
terms of $M$ and $\Lambda$.
The least arbitrary way to do this is to relate the cut-off to
the actual shift of the black hole horizon
with respect to the Schwarzschild-radius $2M$ which has previously
determined to be of the relative order $\frac{\Lambda}{3} (2M)^{2}$.
Thus, we shall subject the choice of $L$ in (53) to a condition of the type
\be
L=r_{b}+l_{c},~~~l_{c} \approx \frac{\Lambda}{3} (2M)^{3}.
\ee
Using the equations (51) and (52) one can show that 
$S_\mu^\nu$ takes the form (in $t,r^*$ coordinates)
\be
S_\mu^\nu(r)=\left(\matrix{S^\alpha_\alpha(r)-\Omega^{-1}(r)H(r)&0\cr
0&\Omega^{-1}(r)H(r)~\cr}\right)+
\Omega^{-1}(r)\left(\matrix{-\beta&-\alpha\cr
\alpha&\beta\cr}\right).
\ee
Now, defining $Q=\alpha+\beta$ and $K=\alpha$, the tensor $S_{\mu}^{\nu}$
takes the form
\be
S_{\mu}^{\nu}=S^{(r)\nu}_{\mu}+S^{(eq)\nu}_{\mu}
\ee
with
\be
S_\mu^{(r)\nu}=\left(\matrix{S_\alpha^\alpha(r)-\Omega^{-1}(r)H(r)&0\cr
0&\Omega^{-1}(r)H(r)\cr}\right)+K\Omega^{-1}(r)\left(\matrix{1&-1\cr
1&-1\cr}\right)
\ee
and
\be
S_{\mu}^{(eq)\nu}=Q\Omega^{-1}\left(\matrix{-1&0\cr
0&1\cr}\right).
\ee
Both tensors in (57) satisfies the conservation law. 
Note that, only $S_{\mu}^{(r)\nu}$
has off-diagonal (flux) components.\\
Now we should determine the constants Q and K. For the determination of
$Q$ we require the regularity of $S_{\mu}^{\nu}$ at
the black hole horizon in a coordinate system which is regular there.
This results in a relation, [Appendix.B]
\be
Q+H(r)\rightarrow 0,~~~as~~~ r \rightarrow r_{b}
\ee
which together with (53) implies 
\be
Q={1/2}\int^{L}_{r_{B}} 
S_\alpha^\alpha(r^\prime){d\over{dr^\prime}}\Omega(r^\prime)dr^\prime.
\ee
Using (54), the approximate value (we neglect terms of higher orders 
in $\Lambda$)   
of this integral can be found to be
\be
Q\approx{\Lambda\over{72\pi}}
\ee
from which one infers that in quasi-flat regions of space-time 
$r \approx r_{q.f}$ where
\be
r_b << r_{q.f}<< r_c,~~~  \Omega(r_{q.f})\approx 1, 
\ee
the tensor $S_\mu^{(eq)\nu}$ in (57) describes an 
equilibrium gas with a temperature $T_c=1/2\pi\sqrt{\Lambda/3}$. 
This follows if one compares the tensor $S_{\mu}^{(eq)\nu}$ with the stress
tensor of an equilibrium gas, namely 
\be
{\pi\over{12}}(kT)^2\left(\matrix{-2&0\cr
0&2\cr}\right).
\ee
The equilibrium temperature $T_c$
corresponds to the leading-order estimate of the temperature of 
the cosmological
event horizon [14].\\
We proceed now to describe the radiation temperature of the black hole.
In present case an outwards flux of 
thermal radiation in quasi-flat regions can be described by the stress tensor
\be
{\pi\over{12}}(kT)^2\left(\matrix{-1&-1\cr
1&1\cr}\right)
\ee
where $T$ is the temperature. For such a stress tensor the energy density
and flux are numerically equal. This latter condition if applied 
in (57) to the tensor $S_\mu^{(r)\nu}$, leads in quasi-flat regions to
the relation 
\be
K={1\over2}\{H(r_{q.f})-S_{\alpha}^{(r)\alpha}(r_{q.f})\}
\ee
in which $H(r_{q.f})={\pi\over6}(8\pi M)^{-2}+O(\Lambda)$, 
as may be verified 
from (53) by a simple calculation.
Therefore $S_\mu^{(r)\nu}$ takes in quasi-flat regions the form
\be
S_\mu^{(r)\nu}(r \to r_{q.f})={\pi\over{12}}(8\pi M)^{-2}
\left(\matrix{-1&-1\cr
1&1\cr}\right)+O(\Lambda)
\ee
from which one infers that $S_\mu^{(r)\nu}$ describes an outwards 
radiation  with the temperature 
\be
T_b=(8\pi M)^{-1}+O(\Lambda M).
\ee
The term $O(\Lambda M)$ is a correction term to the  
Hawking temperature $T_{H}=(8\pi M)^{-1}$ [15] which is the temperature of
the hole in the absence of the cosmological event horizon. 
In terms of the cut-off 
length $l_{c}$
the correction term takes the form $O(l_{c}/M^{2})$. Thus the correction
to the Hawking temperature is a term of the relative order 
$l_{c}/M$. In our case this makes no significant difference for 
thermal predictions because our assumption $\Lambda M^{2}<<1$ means
that the cut-off $l_{c}$ is much smaller than the Schwarzschild-radius
$2M$.

\section{Concluding remarks}

We have seen that 
the existence of a cut-off excluding the contribution
of the anomalous trace to the stress tensor in a neighbourhood of the black
hole horizon
can be connected
to the equilibrium temperature of 
a background heat bath of the cosmological event horizon.
For the corresponding temperature we have found an estimate in terms of the
contribution of the anomalous trace close to the black hole 
horizon, see (61) and (62).
It is important to note that, while the latter contribution seems
to be unphysical with respect to the radiation temperature 
coming from the black hole at late times, 
it does determine the leading order estimate of the equilibrium temperature.
Is there any justification for regarding the contribution of the 
anomalous trace close
to the black hole horizon as
physical with respect to the equilibrium
temperature? We emphasize the distinct character of the equilibrium
temperature as compared to the radiation temperature.
The former
is not expected to be  
sensitive to the outwards signals at late times 
coming from the direction of 
the black hole, so dissatisfaction 
with the role of the infinite gravitational 
redshift at the black hole horizon may not be expressed in this case. \\\\
{\large {\bf Acknowledgment\\}} 
One of us (H. G.) is indebted for advise and other matters in his M.Sc Thesis.
\vskip 1.0cm
{\large {\bf Appendix . A\\}}
Let $n$ be an arbitrary real number.
A tensor $H^{\alpha\beta...}_{\gamma\delta...}$ is said to be of
the covariant order $\Gamma^{n}$ with respect to some
parameter $\Gamma$, if 
$\Gamma^{-n}H^{\alpha\beta...}_{\gamma\delta...}$ can be factrorized
in the metric tensor $g_{\mu\nu}$. 
We shall denote such a situation by
$H^{\alpha\beta...}_{\gamma\delta...}=O(\Gamma^{n})$.(For scalars the
usual meaning is understood). For simplicity the attribute "covariant"
has been suppressed throughout the paper.
From (28) one
gets \be
R=4\Lambda=O(\Lambda)
\ee
and
\be
R^{\mu\nu}=R_\lambda^{\mu\lambda\nu}=
g_{\lambda\gamma}R^{\mu\lambda\nu\gamma}=\Lambda g^{\mu\nu}=O(\Lambda).
\ee
From the last equation it follows
\be
R^{\mu\lambda\nu\gamma}=\Lambda g^{\mu\lambda}g^{\gamma\nu}=O(\Lambda).
\ee
We also find 
\be
v_1(x)=lim_{x^\prime \to x}V_1(x,x^\prime)={1\over720}
\{\Box R-R_{\alpha\beta}
R^{\alpha\beta}+
R_{\alpha\beta\gamma\lambda}
R^{\alpha\beta\gamma\lambda}\}= O(\Lambda^2).
\ee
\vskip 1.0cm
{\large {\bf Appendix . B\\}}
An analysis similar to that presented in [1] for the Schwarzschild-metric 
shows that $S_\mu^\nu,$ as measured in a local Kruskal coordinate system 
at black hole horizon, will be finite if $S_{vv},$ and $S_t^t+S_{r^*}^{r^*}$ 
are finite as $r \to r_b$ and
\be
lim_{r \to r_b}(r-r_b)^{-2}|S_{uu}|<\infty
\ee
where $u$ and $v$ are null coordinates. We find easily
\be
S_{uu}={1\over4}(S_{tt}+S_{r^*r^*}-2S_{tr^*}).
\ee
Using (57)-(59), this gives
\be
S_{uu}={1\over2}\{H(r)+Q-{1\over2}{\Omega(r)S_\alpha^\alpha(r)}\}.
\ee
Therefore, the condition (73) is equivalent to 
\be
H(r)+Q \to 0~~~ as ~~~r \to r_b
\ee
\vskip 1.0cm
\newpage
{\large {\bf References\\}}\\
\begin{tabular}{r p{15cm}}

1. & S.M.Christensen and S.A.Fulling, Phys. Rev. D15, 2088 (1977).\\
2. & Quantum fields in curved space. Birrell $\&$ Davies, chambridge 
     (1982).\\
3. & Adler,s., Lieberman.J., NG, Y.J, Ann. Phys. 106, 279 (1977).\\
4. & R.M.Wald, Commun. Math. Phys, 54, 1-19, (1977).\\
5. & R.M.Wald, Phys. Rev. D17. N6, 1477 (1978).\\
6. & Denis Bernard and Antoine Folacci, Phys. Rev. D34. N8, 2286 (1986).\\
7. & Jacobson Th, Phys. Rev D44. N6, 1731 (1991).\\
8. & Salehi H,Class. Quantum Grav. 10, 595, (1993).\\
9. & Jacobson Th, Phy. Rev, D48, 2, 728, (1993).\\
10. & Christensen.S.M.: Phys. Rev. D14, 2490 (1976).\\
11. & Haag R, Narnhofer H and Stein U, Commun. Math. Phys. 94, 219, (1984).\\ 
12. & M.R.Brown, J. Math. Phys. 25 (1), January (1984).\\
13. & S.Deser,M.J.Duff, and C.J.Isham, Nucl. Phys. B111, 45 (1976).\\
14. & G.W.Gibbons and S.W.Hawking, Phys. Rev. D15, N10, 1738 (1977).\\
15. & S.W.Hawking, Commun. Math. Phys, 43, 199 (1975).\\
\end{tabular}

\end{document}